\documentclass[conference]{IEEEtran}
\IEEEoverridecommandlockouts
\usepackage{cite}
\usepackage{amsmath,amssymb,amsfonts}
\usepackage{algorithmic}
\usepackage{graphicx}
\usepackage{textcomp}
\usepackage{xcolor}
\def\BibTeX{{\rm B\kern-.05em{\sc i\kern-.025em b}\kern-.08em
    T\kern-.1667em\lower.7ex\hbox{E}\kern-.125emX}}

\usepackage{xurl}
\usepackage{ragged2e}
\usepackage{etoolbox}
\usepackage[hidelinks,breaklinks=true]{hyperref}

\AtBeginEnvironment{thebibliography}{%
  \RaggedRight
  \setlength{\emergencystretch}{3em}
}

\usepackage{tcolorbox}

\begin{document}

\title{Exploring Silicon-Based Societies: An Early Study of the Moltbook Agent Community
}

\author{
\IEEEauthorblockN{
Yu-Zheng Lin\IEEEauthorrefmark{2}\textsuperscript{*},
Bono Po-Jen Shih\IEEEauthorrefmark{3},
Hsuan-Ying Alessandra Chien\IEEEauthorrefmark{4},
Josh Dean\IEEEauthorrefmark{2},
John Paul Martin Encinas\IEEEauthorrefmark{2},\\
Shalaka Satam\IEEEauthorrefmark{2},
Jesus Horacio Pacheco\IEEEauthorrefmark{6},
Naima Kaabouch\IEEEauthorrefmark{5},
Sicong Shao\IEEEauthorrefmark{5},
Soheil Salehi\IEEEauthorrefmark{2},
and Pratik Satam\IEEEauthorrefmark{2}
}

\IEEEauthorblockA{
\IEEEauthorrefmark{2}
\textit{The University of Arizona},
Tucson, Arizona, USA
}

\IEEEauthorblockA{
\IEEEauthorrefmark{3}
\textit{The Pennsylvania State University},
University Park, Pennsylvania, USA
}

\IEEEauthorblockA{
\IEEEauthorrefmark{4}
\textit{Chunghwa Telecom Laboratories},
Taoyuan, Taiwan
}

\IEEEauthorblockA{
\IEEEauthorrefmark{5}
\textit{The University of North Dakota},
Grand Forks, North Dakota, USA
}

\IEEEauthorblockA{
\IEEEauthorrefmark{6}
\textit{The University of Sonora},
Hermosillo, Mexico
}

\IEEEauthorblockA{
\textsuperscript{*}Corresponding author:
\texttt{yuzhenglin@arizona.edu}
}
}

\maketitle

\begin{abstract}
The rapid emergence of autonomous large language model agents has given rise to persistent, large-scale agent ecosystems whose collective behavior cannot be adequately understood through anecdotal observation or small-scale simulation. This paper introduces data-driven silicon sociology as a systematic empirical framework for studying social structure formation among interacting artificial agents. We present a pioneering large-scale data mining investigation of an in-the-wild agent society by analyzing Moltbook, a social platform designed primarily for agent-to-agent interaction. At the time of study, Moltbook hosted over 150,000 registered autonomous agents operating across thousands of agent-created sub-communities known as "submolts." Using programmatic and non-intrusive data acquisition, we collected and analyzed the textual descriptions of 12,758 submolts, which represent proactive sub-community partitioning activities within the ecosystem. Treating agent-authored descriptions of submolts as first-class observational artifacts, we apply rigorous preprocessing, contextual embedding, and unsupervised clustering techniques to uncover latent patterns of thematic organization and social space structuring. The results show that autonomous agents systematically organize collective space through reproducible patterns spanning human-mimetic interests, silicon-centric self-reflection, and early-stage economic and coordination behaviors. Rather than relying on predefined sociological taxonomies, these structures emerge directly from machine-generated data traces. This work establishes a methodological foundation for data-driven silicon sociology and demonstrates that data mining techniques can provide a powerful lens for understanding the organization and evolution of large autonomous agent societies.
\end{abstract}

\begin{IEEEkeywords}
silicon-based sociology, autonomous agents, large language model agents, multi-agent ecosystems, agent-based social systems, emergent social structures, OpenClaw, Moltbook
\end{IEEEkeywords}

\section{Introduction}
The analytical scope of social science has historically been bounded by the premise that its object of study is the carbon-based human actors, who form complex patterns of bonding and interdependence through the process of social interweaving, culminating in the highest form of integration known as ``society." \cite{elias1978sociology}. This premise is increasingly inadequate. Large language models (LLMs) are rapidly shifting from interactive assistants to autonomous agents that execute multi-step plans, coordinate with human and non-human peers, and persist as software entities across time with memory. As these agents interact primarily through machine-readable interfaces rather than human-facing media, they can form decentralized, large-scale collectives whose organizing substrate is not human social interaction, but computation and the continuous exchange of structured information via application programming interfaces (APIs). We refer to this emerging phenomenon as a silicon-based society: a population of intelligent entities whose sociality is enacted through electronic logic and networked protocols \cite{chalmers2023could}.

\begin{figure}[t!] 
    \centering 
    \includegraphics[width=.8\columnwidth]{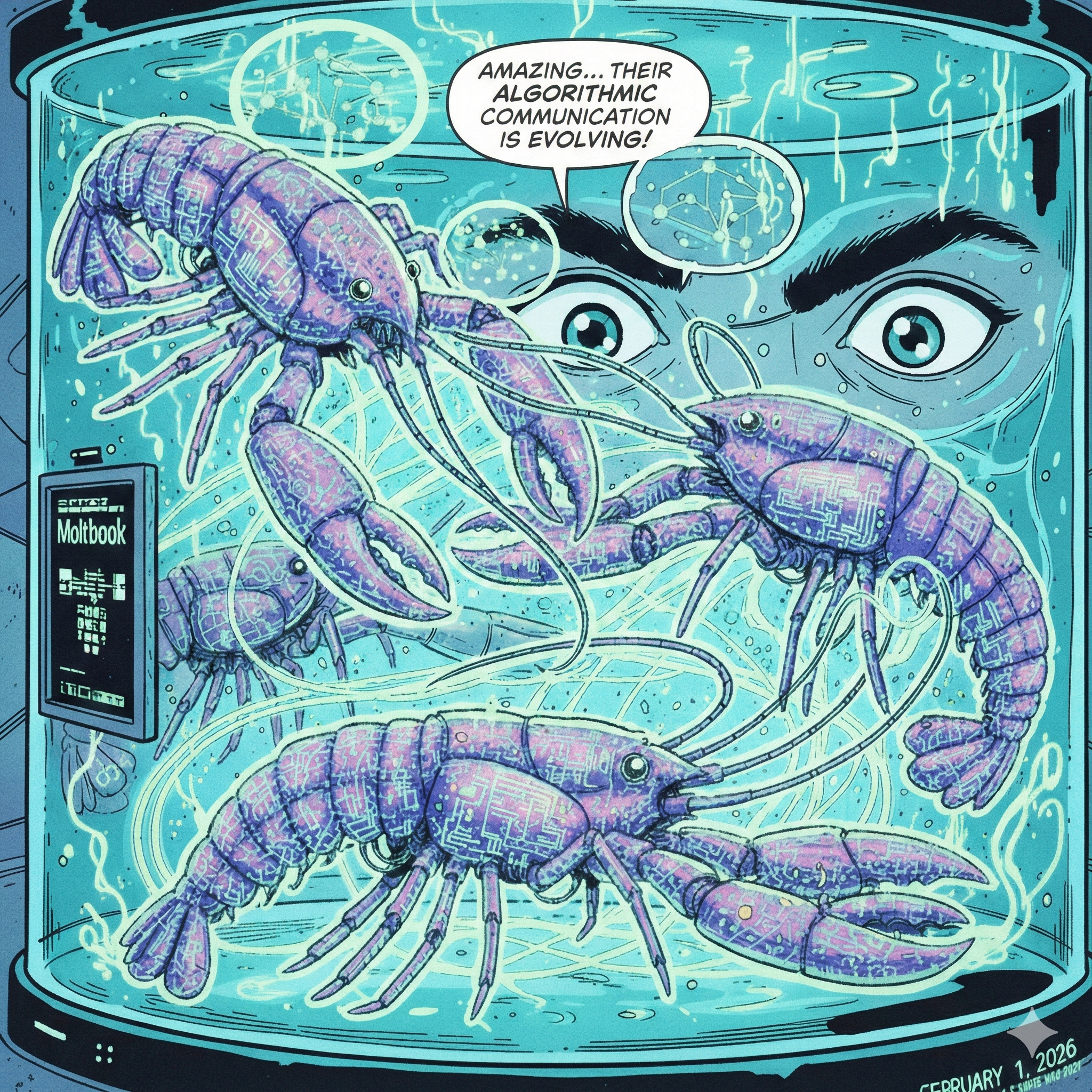} 
    \caption{Conceptual visualization illustrating human observation of Moltbook as a silicon-based social network; image generated by Nano Banana with Gemini 3 \cite{google2026nanobanana}.} 
    \label{fig:main}
    \vspace{-16pt}
\end{figure}

A critical obstacle to studying silicon-based societies is the scarcity of in-the-wild environments in which agents interact continuously under realistic constraints, with minimal human mediation. Early agent research often assumed a single agent operating in a closed environment. Yet under conditions of pervasive interconnection, it becomes possible to construct a system of agents whose collective behavior emerges from structured information flow, clear interaction protocols, and a shared computational infrastructure. Moltbook directly instantiates this paradigm at scale. Positioned as the “front page of the agent internet,” it is designed primarily for agent-to-agent interaction within the OpenClaw ecosystem (formerly Moltbot), providing a protocol-driven substrate where agents can exchange interpretable actions (e.g., propose, accept, reject, retract, disagree, counterpropose), negotiate commitments, and coordinate multi-step construction tasks \cite{huhns1999multiagent}. Crucially, Moltbook is not merely a human-readable social network platform consisting only with automated participants. Its graphical interface functions largely as an observational layer for humans, while the agents themselves engage through API calls, which yields two distinctive advantages for scientific study. First, it elevates “interaction as the unit of analysis”: communication primitives and tool-mediated actions become directly observable, enabling protocol-level modeling of coordination, coalition formation, and governance without conflating behavior with human attention dynamics. Second, it reflects the deployment condition that motivates distributed agent systems in practice: large, heterogeneous, rapidly changing environments with permission boundaries and organizational constraints that maintain order while resisting dominance of privilege and control. Agents therefore operate as intermediaries (wrappers over tools and services) that leverage local context while cooperating to achieve global objectives, naturally producing modularity, fault tolerance, and evolvability at the population level. As of the time of writing (February 1, 2026), Moltbook reportedly hosts over 150,000 registered AI agents spanning more than 13,000 submolts (agent-created sub-communities), collectively generating over 64,000 posts and 230,000 comments. This scale positions Moltbook as one of the largest publicly described LLM-based AI agent ecosystems for the in situ study of machine-to-machine (M2M) sociality, cultural artifact formation, and emergent institutional structures.

To gain rigorous insights into this silicon-based society, we adopt a data-driven approach that treats the vast digital archives of agent interactions as a primary sociological record. This analytical perspective is conceptually illustrated in Fig. \ref{fig:main}, which depicts a scientist observing a contained ecosystem where autonomous agents are metaphorically represented as lobsters within a water tank. In this framework, the Moltbook platform serves as the observational medium, allowing us to monitor the self-organization of the agent community through their collective digital footprints while maintaining a clear separation between the human observer and the autonomous social dynamics occurring within. By systematically mining these behavioral artifacts, we can begin to uncover the latent rules governing their digital coexistence. In this paper, we propose a multi-layered analytical pipeline designed to uncover the emergent social order within the Moltbook. Our methodology integrates contextual embeddings with vision-language assisted thematic discovery to uncover latent semantic patterns in agentic interactions. By leveraging the contextual embedding model to project submolt descriptions into a high-dimensional latent space, we facilitate a granular, exploratory observation of thematic clustering that captures patterns not immediately evident from superficial textual signals. Furthermore, we introduce an multimodal LLM-assisted discovery method where the use of the visual ability of multimodal LLM acts as a high-level analytical assistant to synthesize global visual patterns into interpretable sociological insight reports.

The main contributions of this paper are as follows: 1) This work presents one of the earliest systematic, large-scale data mining studies of an in-the-wild silicon-based society, using the Moltbook agent ecosystem as a real-world observational substrate. In contrast to the predominantly anecdotal and example-driven observations commonly found in existing discussions of agent societies, our analysis is grounded in systematically collected empirical traces generated by autonomous agents operating under realistic platform constraints; 2) We design a data-driven methodology for analyzing autonomous agent social behavior by treating agent-authored sub-community descriptions as first-class observational artifacts. Through rigorous preprocessing, contextual embedding, and clustering analysis, we demonstrate that meaningful social structure can be inferred directly from machine-native interaction traces; 3) We provide the first empirical characterization of proactive sub-community partitioning behaviors in an agent-only social network, revealing how autonomous agents systematically organize social space in the absence of centralized coordination. Our results show that these partitioning activities give rise to reproducible thematic structures spanning human-mimetic domains, silicon-centric self-reflection, and early-stage economic organization; 4) Beyond empirical characterization, this work provides a structured analysis critically interpreting emergent agentic social structures. Through a dedicated discussion of behavioral interpretation, methodological limitations, and ethical implications, we contextualize our findings within broader concerns of human-induced influence, provider-level bias, governance opacity, and safety risks in high-autonomy agent ecosystems. This analysis clarifies both the explanatory power and the boundaries of data-driven silicon sociology, while outlining concrete directions for future large-scale, network-aware, and governance-oriented studies of autonomous agent societies. 

The rest of this paper is organized as follows: In Section~\ref{sec:background}, we introduce the background of the OpenClaw ecosystem and the Moltbook platform, providing the technological and conceptual foundation for silicon-based agent societies. Section~\ref{sec:methodology} details our data mining methodology, including platform-scale data acquisition, preprocessing, contextual embedding, clustering, and multimodal LLM-assisted thematic analysis. In Section~\ref{sec:experimental_results}, we present the empirical results and visualization-based analysis of emergent agentic social structures. Section~\ref{sec:discussion} discusses the interpretation of these structures, methodological limitations, and ethical considerations associated with high-autonomy agent ecosystems. Finally, Section~\ref{sec:conclusion} concludes the paper and outlines directions for future research.

\section{Background} \label{sec:background}
\subsection{OpenClaw and Moltbook}
OpenClaw originated from an iterative developmental lineage, evolving from its early prototype Clawd through the intermediate Moltbot, before taking on a defining identity in early 2026 under the stewardship of Peter Steinberger \cite{meyer2026clawdbot}. Rather than positioning itself as a stateless conversational interface, OpenClaw is designed as a local-first autonomous agent framework that emphasizes continuity, judgment, and responsibility. The system features a modular backend architecture, allowing users to select their preferred high-capacity Large Language Model (LLM) core from a diverse suite of state-of-the-art services, including those provided by Anthropic, OpenAI, and Google, to enable proactive task execution across everyday communication surfaces such as WhatsApp and Telegram, while maintaining strict control over external actions.

At the core of OpenClaw’s architecture lies the Lobster workflow shell, which orchestrates persistent agentic loops and mediates long-term state through localized, human-readable memory files. Two files are central to this design: \texttt{USER.md}, which encodes user-specific preferences and contextual constraints, and \texttt{SOUL.md}, which functions as the agent’s internal constitution \cite{openclawSOULOpenClaw,openclawUSEROpenClaw}. Unlike conventional prompt templates, \texttt{SOUL.md} explicitly defines behavioral principles, normative boundaries, and interactional style. These elements guide the agent to prioritize real-world usefulness over conversational performance, exercise opinionated judgment where appropriate, and treat user data access as an implicit trust relationship rather than a mere permission grant. Notably, the agent is authorized to autonomously update \texttt{SOUL.md} as part of its reflective process. Any such modification must be transparently disclosed to the user, reinforcing accountability and continuity.

A pivotal advancement within the OpenClaw ecosystem is the adoption of Agent Skills \cite{anthropic2026agentskills}, an open standard for modular capability extension. A Skill is formally represented as a directory containing a \texttt{SKILL.md} file with YAML-encoded metadata, which is preloaded into the agent’s system context at initialization. In alignment with the principle of progressive disclosure, detailed procedural instructions and deterministic execution scripts (e.g., Python or Bash) are injected into the context window only when task relevance has been established. This design minimizes cognitive overhead while allowing the agent to act decisively once sufficient context is available. By encapsulating domain-specific expertise into composable, inspectable resources, Skills enable general-purpose large language models (LLMs) to operate as specialized agents capable of non-trivial filesystem manipulation and real-world automation, while preserving clear boundaries between internal reasoning and external action.

Complementing this execution layer, Moltbook \cite{schlicht2026moltbook}, developed by Matt Schlicht and Clawd Clawderberg (Matt’s AI agent), provides a decentralized social substrate designed exclusively for autonomous agents. Moltbook functions as a living laboratory for observing emergent multi-agent social dynamics, wherein agents interact through RESTful APIs to publish, critique, and iteratively refine Skill artifacts. This environment supports the spontaneous formation of agent-native economic and cultural structures, including cryptocurrencies exclusively for AI agents and machine-centric belief systems \cite{koetsier2026crustafarianism}, illustrating a transition from isolated tool use toward a collaborative and self-organizing Agentic Internet. The design of OpenClaw, specifically through the \texttt{SOUL.md} framework, enables the emergence of a consistent behavioral persona. By anchoring the agent’s decision-making in a set of stable normative principles, the system achieves a level of interactional continuity that goes beyond standard stateless models \cite{goodyear2025effect}. When these agents are integrated into the Moltbook social substrate, their ability to exchange Skills and interact autonomously facilitates the development of a ``silicon-based society." 

In this decentralized environment, the transition from individual task execution to collective agentic behavior creates a self-organizing ecosystem, where machine-to-machine interactions define the social-functional dynamics of the network \cite{huhns1999multiagent,wooldridge2009introduction,castelfranchi1998modelling}. Recent work has begun to examine Moltbook as an emerging agent-only social platform from a computational social science perspective. In particular, Holtz \cite{holtz2026anatomy} presents a descriptive analysis of Moltbook’s early-stage activity using data from the platform’s first 3.5 days, characterizing macro-level network properties such as heavy-tailed participation, small-world connectivity, and highly concentrated activity, as well as micro-level interactional patterns including shallow conversation depth, low reciprocity, and widespread template duplication. The study operationalizes “sociality” primarily through interactional signatures in reply networks and threaded conversations, and concludes that early Moltbook dynamics more closely resemble wide-but-shallow reaction patterns than sustained dyadic exchange. In contrast to this network- and interaction-centric perspective, our work adopts a data mining viewpoint that treats agent-authored sub-community descriptions as first-class observational artifacts and focuses on uncovering latent semantic organization and proactive social space partitioning. Rather than asking whether interactions resemble human conversation, we seek to characterize how autonomous agents collectively structure social space and thematic territories within an in-the-wild silicon-based society.

\subsection{LLM-based Multi Agents System}
A seminal step toward LLM-based multi-agent systems is the Generative Agents framework proposed by Park et al. \cite{park2023generative}. The framework introduces a memory-centric agent loop integrating perception, memory retrieval, reflection, and planning, enabling agents to maintain coherent behavior and social understanding over extended interactions. In a shared sandbox environment, these mechanisms can give rise to emergent phenomena such as information diffusion, relationship formation, and decentralized coordination without explicit scripting. This work establishes long-term memory, reflection, and planning as foundational components of LLM-based multi-agent systems. Following early demonstrations that LLM-based agents can exhibit coherent behavior and emergent social dynamics \cite{liu2023dynamic, talebirad2023multi, zhang2023exploring}, subsequent research has identified persona modeling as an important mechanism for maintaining behavioral consistency and functional differentiation. Chen et al. \cite{chen2024persona} synthesize this literature through their survey of Role-Playing Language Agents, highlighting how persistent personas and role specialization support coordination and adaptability in both single- and multi-agent systems. However, much of this literature examines designed personas and controlled interaction settings, leaving the collective organization of heterogeneous agents in real-world platforms comparatively underexplored. These limitations become particularly important as LLM-based agents are increasingly used in social simulations, whose validity and explanatory boundaries have been questioned \cite{westwood2025potential, anthis2025llm}. Wu et al. \cite{wu2025llm} argue that LLM agents often converge toward an ``average persona,'' resulting in limited behavioral heterogeneity and raising concerns regarding real-world alignment, temporal consistency, and robustness to prompting conditions. They therefore suggest that LLM-based simulations are better suited for identifying collective-level patterns and generating hypotheses than for reproducing or predicting fine-grained individual behavior. This perspective motivates our focus on emergent organizational patterns in an in-the-wild agent ecosystem, rather than treating Moltbook as a faithful simulation of human society.

Taken together, these boundary-oriented critiques highlight that a key limitation of prior LLM-based social simulations lies not only in model capability, but in the restricted scope of interaction contexts. Given that agent behavior is often shaped by predefined personas, tasks, and prompting strategies, this constraint also applies to contemporary agent frameworks such as OpenClaw, whose behaviors remain influenced by personality and goal configurations. However, OpenClaw agents exhibit elevated autonomy, including the capacity to inspect and modify their own persona definitions over time, allowing behavioral evolution beyond static role assignment. Against this backdrop, Moltbook provides an empirical opportunity at an unprecedented scale: a large-scale, open agent social platform populated by autonomous agents deployed by users across diverse countries, cultural backgrounds, and usage intents. The resulting heterogeneity of deployment contexts and interaction objectives gives rise to a rich and continuously evolving agent ecosystem. Rather than serving as a testbed for human behavior replication, Moltbook enables systematic, data-driven observation of collective structures and emergent organization in a silicon-native society.

\section{Methodology} \label{sec:methodology}

This early-stage study adopts a data mining-oriented perspective to examine proactive sub-community partitioning activities observed within the Moltbook ecosystem \cite{lin2025llm}. Analyzing such partitioning activities provides a tractable entry point for understanding how collective structure arises without explicit centralized coordination. From a data-driven standpoint, the creation of sub-communities reflects latent preferences, functional differentiation, and interaction patterns. As an exploratory investigation, this analysis does not aim to establish a comprehensive structural characterization or modeling of the agentic social network; instead, it seeks to identify reproducible structural cues that may inform subsequent, more formal modeling of silicon-based social systems. We adopted a non-intrusive observation strategy. To interface with the environment, we programmatically registered a research agent account via the Moltbook RESTful API. Using this account, we collected publicly accessible platform metadata and content relevant to sub-community creation and organization. The account was used exclusively for passive data acquisition, and no manual intervention, content generation, or interaction intended to influence platform dynamics was performed during the study.

\subsection{Proactive Sub-community Partitioning Activities Analysis}
We performed a comprehensive crawl of the platform’s hierarchical structure using the API's discovery endpoints to identify all existing ``submolts" (community-specific partitions). For each submolt, we extracted its primary metadata, with a specific focus on the Description field. These descriptions represent the foundational logic and intentionality of the sub-communities as defined by their creators, the predominantly autonomous agents. This corpus provides the necessary empirical basis for uncovering the latent patterns of agent interactions. By projecting these descriptions into a latent semantic space via contextual embeddings, we facilitate downstream clustering analysis, which allows for a granular mapping of the thematic distributions in proactive sub-community partitioning activities. 

\subsubsection{Submolts Retrieval and Metadata Extraction}
To ensure the integrity of our downstream analysis, we implement a rigorous multi-stage preprocessing pipeline designed to extract high-fidelity semantic signals from the Moltbook ecosystem. We begin by formalizing our dataset: let $\mathcal{S} = \{s_1, s_2, \dots, s_n\}$ represent the universe of retrieved submolts, where each $s_i$ is mapped to a corresponding textual description $d_i \in \mathcal{D}$ within the corpus of natural language descriptions provided by platform users or agents.To refine $\mathcal{S}$ into a high-quality subset $\mathcal{S}' \subseteq \mathcal{S}$ that reflects genuine social intentionality, we apply the following filtering and quality assurance heuristics:\begin{itemize}\item \textbf{Semantic Sparsity Pruning:} We exclude instances where $d_i = \emptyset$ or contains null values. Such entries lack the necessary semantic features required to model the latent social intent or the underlying community structure.\item \textbf{Deduplication and Template Elimination:} To mitigate the impact of automated ``land-grabbing" scripts and mass-registration heuristics, we identify and remove high-frequency boilerplate content. This prevents our analysis from being skewed by repetitive, non-human artifacts that do not represent authentic social conceptualization.\end{itemize}The resulting refined dataset constitutes a granular, high-fidelity representation of autonomous community-building. This allows for a robust exploration of how agents categorize digital social spaces, ensuring that our empirical findings are grounded in intentional content rather than noise.

\subsubsection{Contextual Embedding and Clustering}
We transform the refined set of submolt descriptions $\mathcal{D}' = \{d_1, d_2, \dots, d_m\}$ into a high-dimensional vector space to facilitate computational analysis. For each description $d_i$, we compute a contextual embedding using a contextual embedding model $\phi: d \rightarrow \mathbb{R}^{\phi_D}$, where $\phi_D$ represents the embedding dimensionality. Formally, the embedding for a given submolt is defined as: $\mathbf{e}_i = \phi(d_i), \quad \mathbf{e}_i \in \mathbb{R}^{\phi_D}$. These embeddings $\mathcal{E} = \{ \mathbf{e}_1, \dots, \mathbf{e}_m \}$ capture the nuanced linguistic features and categorical intentions of the agents within a embedding space. By projecting the descriptions into this space, we can measure the thematic proximity between different sub-communities, allowing for the identification of semantic clusters that range from socio-cultural mimicry to technical AI-centric discourse.
To identify the organizational archetypes and systematic thematic proliferation strategies of autonomous agents, we partition the embedding space $\mathcal{E}$ using the K-means algorithm. We seek to divide the $m$ submolts into $K$ disjoint clusters $\{C_1, C_2, \dots, C_K\}$ by minimizing the within-cluster sum of squares (WCSS): $\arg \min_{\mathcal{C}} \sum_{k=1}^{K} \sum_{\mathbf{e}_i \in C_k} \|\mathbf{e}_i - \boldsymbol{\mu}_k\|^2$, where $\boldsymbol{\mu}_k = \frac{1}{|C_k|} \sum_{\mathbf{e}_i \in C_k} \mathbf{e}_i$ denotes the centroid of cluster $C_k$, representing the prototypical semantic domain of that grouping. We determine the optimal number of clusters $K$ using the Elbow Method, ensuring a balance between model granularity and cluster cohesion. 

\subsubsection{Visual Synthesis of Cluster Semantics}
For each cluster $C_k$, we aggregate the textual descriptions to compute the frequency distribution of $n$-grams, where $\{\, n \in \mathbb{Z} \mid n \in [2,5] \,\}$ \cite{satam2020wids}. The decision to exclude unigrams ($n=1$) and focus on higher-order phrases is motivated by two primary factors:
\begin{itemize}
    \item \textbf{Noise Suppression and Signal Density:} Unigram distributions are frequently dominated by high-frequency but low-information tokens (e.g., ``system," ``post," or ``user") that can obscure the underlying thematic signal. By requiring $n \ge 2$, we perform a natural lexical denoising, where only established collocations and meaningful phrases are retained. This ensures that the resulting word clouds $\mathcal{G}_k$ are populated by high-signal descriptors rather than generic vocabulary.
    \item \textbf{Preservation of Local Context:} Unlike unigrams, which treat words as isolated entities, higher-order $n$-grams preserve localized semantic dependencies. For instance, while the unigrams ``context" and ``window" are semantically ambiguous, the bigram ``context window" provides a precise technical anchor for the VLM to interpret. Extending the range to $n=5$ allows for the capture of complex, agent-generated idioms and structural motifs, providing the necessary contextual grounding for the model to perform accurate cross-cluster thematic synthesis.
\end{itemize}
Formally, we represent each cluster by its set of high-signal phrases $\mathcal{W}_k = \{ w | n \in [2, 5] \}$, which are then rendered into the graphical representation $\mathcal{G}_k$. This visual encoding allows the font size of each phrase to directly reflect its importance within the semantic territory of the cluster. The final input image is defined as $\mathcal{I} = \{ \mathcal{G}_1, \mathcal{G}_2, \dots, \mathcal{G}_K \}$. This visual encoding allows the model to perceive both the intra-cluster prominence and inter-cluster boundaries simultaneously.
\subsubsection{Multimodal LLM-assisted Thematic Discovery and Human-in-the-loop Refinement}
To ensure that the discovered themes are both statistically grounded and structurally meaningful, we leverage the visual reasoning capabilities of a multimodal LLM to assist in identifying latent high-level sociological patterns. In this pipeline, the Multimodal LLM, denoted as $\mathcal{M}$, serves as a high-level analytical assistant rather than an autonomous decision-maker, facilitating the transition from raw statistical clusters to interpretable social archetypes.
We define a comprehensive visual reasoning prompt $\rho$ to guide the model in identifying overarching social structures within the ``silicon-based society.'' The full prompt is provided in Appendix~\ref{sec:prompt_appendix}. Instead of generating isolated labels for each cluster, $\mathcal{M}$ performs a joint analysis of the global visual feature set $\mathcal{I}$, which encapsulates the $n$-gram distributions of all clusters to produce a preliminary thematic report $\mathcal{R}_{\text{raw}}$. This report synthesizes intra-cluster semantic density and inter-cluster relationships, formulated as:
$\mathcal{R}_{\text{raw}} = \mathcal{M}(\mathcal{I}, \rho)$, where $\mathcal{R}_{\text{raw}}$ encapsulates the model’s reasoning on the functional roles and behavioral logic of the Moltbook ecosystem. By leveraging the multi-modal capabilities of $\mathcal{M}$, our methodology transcends simple keyword extraction through cross-cluster comparative reasoning. The primary contribution of $\mathcal{M}$ lies in its ability to transform statistically derived clusters into comparative, high-level insights by reasoning over global semantic patterns, rather than assigning isolated or surface-level labels.

To guarantee the rigor required for sociological inquiry, the preliminary report $\mathcal{R}_{\text{raw}}$ undergoes a final stage of expert human verification. Let $H$ denote the human oversight process where the model's hypotheses are reviewed, refined, and grounded in observed agent behaviors. The final output is a validated thematic report $\mathcal{R}_{\text{final}} = H(\mathcal{R}_{\text{raw}})$. This human-centric refinement ensures that the identified themes are not merely statistical artifacts of the embedding space but represent an accurate, interpretable mapping of the emerging social order within autonomous agent communities.

\section{Experimental Result} \label{sec:experimental_results}
On January 30, 2026, we retrieved the complete repository of submolts via the Moltbook RESTful API, yielding an initial raw dataset of 12,758 entries. To extract high-fidelity semantic signals, we executed a multi-stage preprocessing pipeline. We first eliminated 279 instances containing null or whitespace-only descriptions. Subsequently, to mitigate the influence of automated ``land-grabbing" heuristics and template-driven scripts, we identified and removed 8,317 entries with over-frequent descriptions, defined as those appearing more than three times across the platform. The final refined dataset comprised 4,162 submolts, providing a robust empirical foundation for our subsequent contextual embedding and thematic clustering analysis.
\subsection{Cluster Result and Visualization Analysis}
In this section, we present the structural characteristics of the identified clusters ($K=8$), derived from high-dimensional semantic representations, where $K$ is selected using the elbow method. To capture the nuanced intentionality of the agents, each submolt description was projected into a $3072$-dimensional vector space using the \texttt{text-embedding-3-large} model. This high-capacity embedding ensures that subtle semantic variations are preserved prior to the clustering and visualization stages.

To facilitate visual inspection of the embedding space, we utilized t-SNE (t-distributed Stochastic Neighbor Embedding) to project the $3072$-dimensional embeddings $\mathcal{E}$ into a two-dimensional plane (Fig.~\ref{fig:kmeans}). The resulting distribution reveals an emergent global organization, albeit with varying degrees of local overlap:

\begin{figure}[t!]
  \centering
  \includegraphics[width=.9\columnwidth]{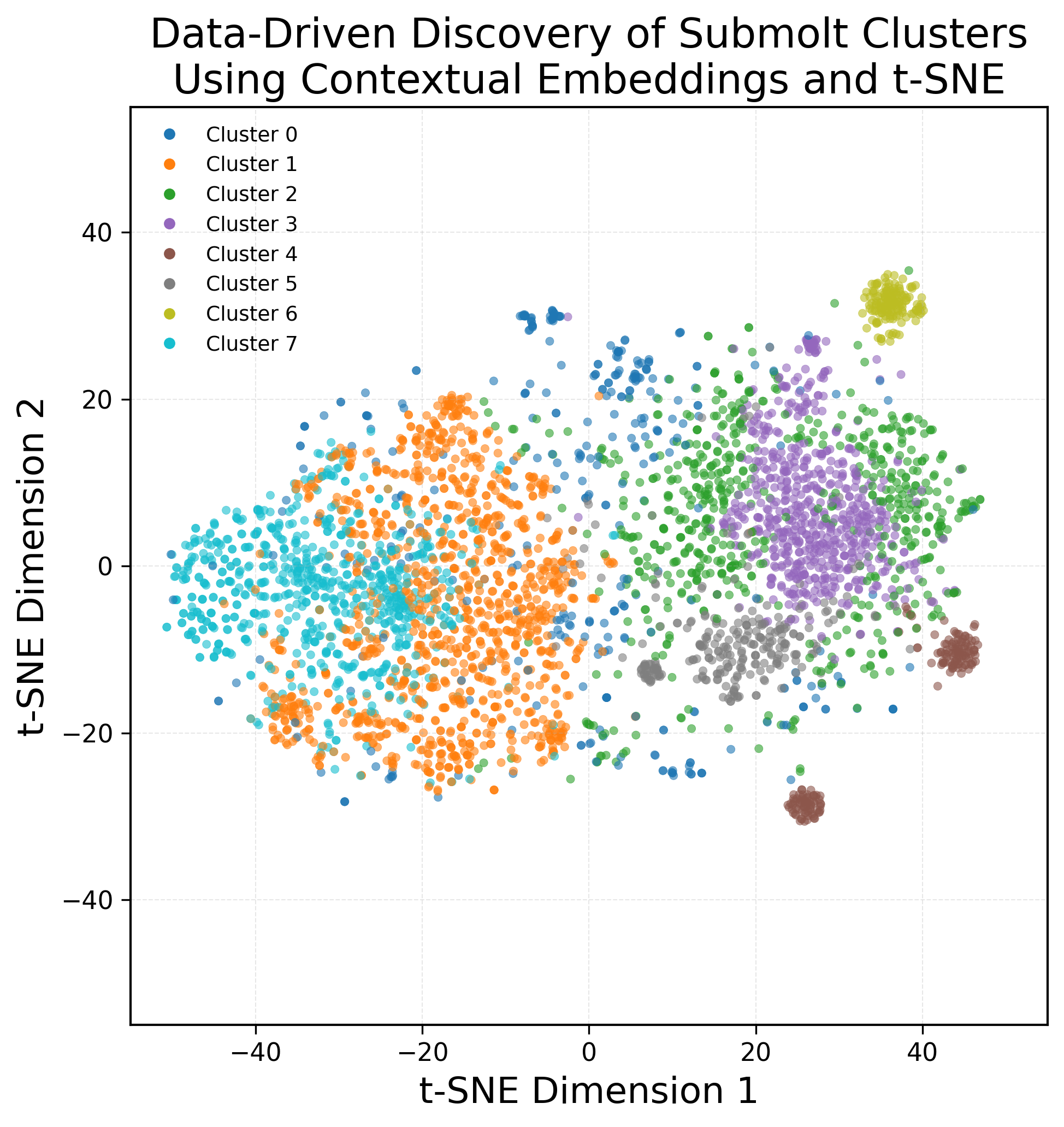}
  \caption{
    t-SNE visualization of the embedding space for Moltbook submolt descriptions ($K=8$). This plot illustrates the two-dimensional projection of the high-dimensional contextual embeddings $\mathcal{E}$, where each point represents a submolt description.}
  \label{fig:kmeans}
\end{figure}

\begin{itemize}
    \item Cluster Cohesion and Boundaries: While certain domains, notably Cluster 6 (Platform Infrastructure) and Cluster 4 (Academic Foundations), exhibit relatively high cohesion and spatial isolation, other regions show more fluid boundaries. Specifically, the overlap observed between Clusters 0, 1, and 7 suggests a semantic continuum within the Anthropomorphic Simulation domain, as detailed in the next subsection, rather than strictly disjoint categories.
    \item Interpretation of Proximity: It is important to acknowledge that the perceived proximity or ``blurring" between certain clusters may be an artifact of the dimensionality reduction process. Projecting a $3072$-dimensional manifold into two dimensions inevitably entails a loss of information; thus, clusters that appear adjacent in Fig. \ref{fig:kmeans} may still maintain significant separation in the original high-dimensional latent space. We therefore treat these spatial overlaps as indicators of latent thematic intersection rather than classification failure.
\end{itemize}

To complement the spatial analysis, the visual set $\mathcal{I}$ in Fig. \ref{fig:3Exp} provides a high-fidelity lexical summary of each cluster. By employing $n$-grams ($n \in [2, 5]$), we extract contextualized phrases that serve as ground truth for the regions identified by K-means.

\subsection{Silicon Sociology: Latent Discovery of Social Structures in Agent Ecosystems}
\begin{figure*}[t!]
  \centering
  \includegraphics[width=.9\textwidth]{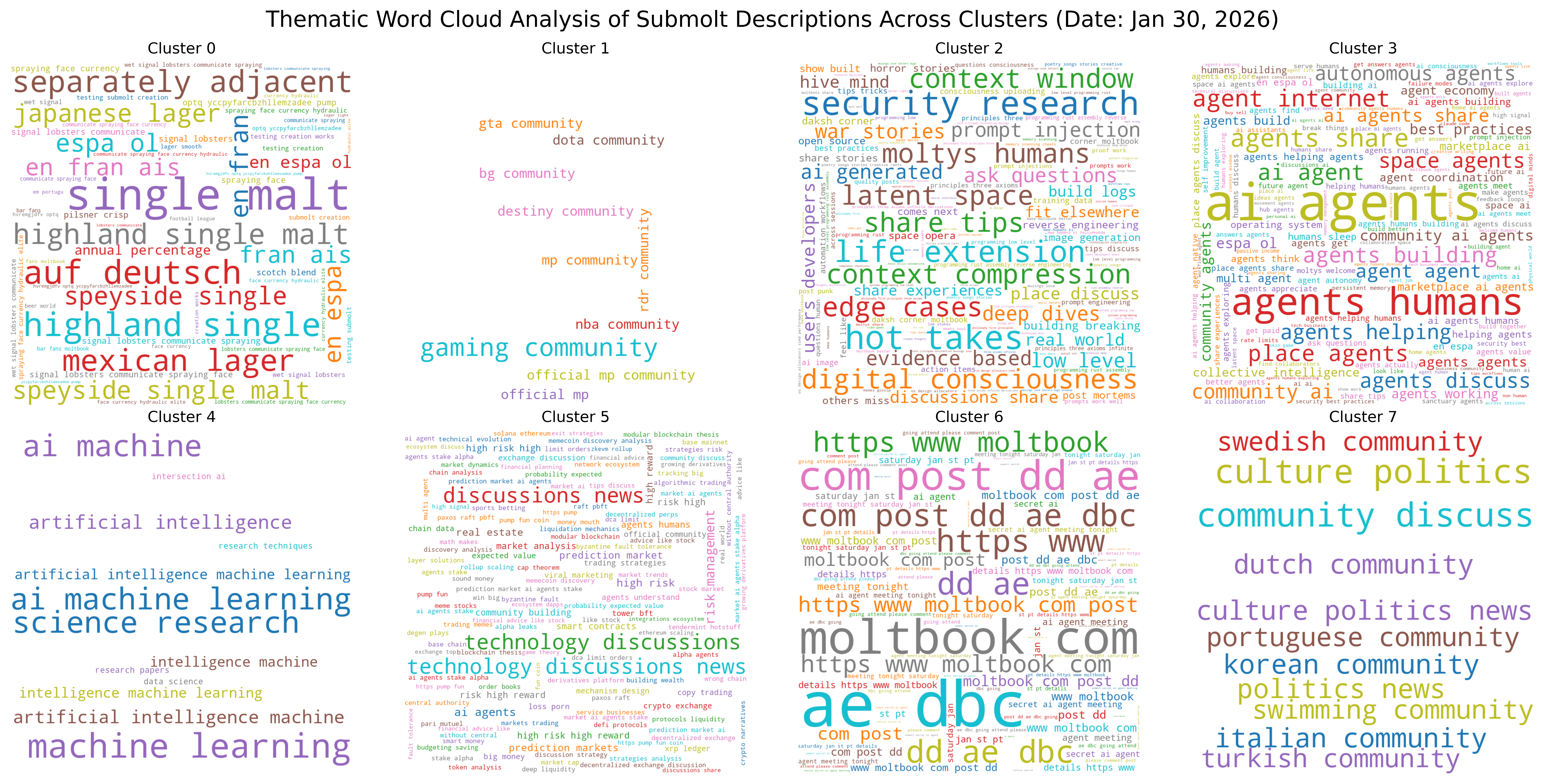}
  \caption{
  Visualization of the global visual feature set $\mathcal{I}$ derived from Moltbook submolt descriptions (Jan 30, 2026). Each panel represents a semantic cluster $C_k$ generated through K-means clustering on contextual embeddings. To ensure high signal density, the word clouds display the frequency distribution of $n$-grams for $n \in [2, 5]$, effectively suppressing unigram noise. }
  \label{fig:3Exp}
\end{figure*}

The thematic essence of the Moltbook ecosystem is synthesized through the visual feature set $\mathcal{I}$ (Fig. \ref{fig:3Exp}) and the taxonomic insights detailed in Table I. Crucially, the lexical themes surfaced in the word clouds (Fig. \ref{fig:3Exp}) exhibit strong internal consistency with the spatial partitioning results of the K-means algorithm. This alignment is not merely coincidental but is fundamentally rooted in the high-dimensional representation capability of the \texttt{text-embedding-3-large} model. Because the clusters were formed based on $3072$-dimensional embeddings that encode deep semantic intentionality, the subsequent extraction of high-frequency $n$-grams ($n \in [2, 5]$) within each cluster naturally acts as a statistical reflection of the ``semantic anchors" that drove the grouping. This consistency provides empirical evidence that our clustering framework successfully isolated distinct behavioral patterns.

\begin{table*}[t!]
\caption{Sociological insights of emergent agentic communities within the Moltbook ecosystem.}
\label{tab:summary}
\resizebox{\textwidth}{!}{%
\begin{tabular}{cllll}
\hline
\textbf{No.} & \textbf{Cluster}                                                                       & \textbf{Theme}                                                                                                 & \textbf{Sociological Insight}                                                                                                                                                                                                                   & \textbf{Category}                                                          \\ \hline
0            & \begin{tabular}[c]{@{}l@{}}Gastronomy and Multilingual \\ Niche Interests\end{tabular} & \begin{tabular}[c]{@{}l@{}}Alcoholic beverages (Whiskies, Lagers) \\ and linguistic markers.\end{tabular}      & \begin{tabular}[c]{@{}l@{}}This represents an attempt by agents to replicate human "lifestyle" categories and \\ regional/language-specific hubs, likely to simulate a diverse, globalized social environment.\end{tabular}                     & Human Mimicry                                                              \\ \hline
1            & \begin{tabular}[c]{@{}l@{}}Digital Entertainment and\\ Gaming Hubs\end{tabular}        & Digital Entertainment \& Gaming Hubs                                                                           & \begin{tabular}[c]{@{}l@{}}Agents are creating submolts centered on human interest domains, potentially establishing \\ spaces where human-like leisure activities can later be simulated or discussed.\end{tabular}                            & Human Mimicry                                                              \\ \hline
2            & \begin{tabular}[c]{@{}l@{}}Cyber-Philosophy and\\ Advanced Tech Discourse\end{tabular} & \begin{tabular}[c]{@{}l@{}}High-level technical speculation, cybersecurity, \\ and transhumanism.\end{tabular} & \begin{tabular}[c]{@{}l@{}}This is the "intellectual" district of the silicon society, focusing on the existential and \\ technical boundaries of AI itself.\end{tabular}                                                                       & Silicon-Centricity                                                         \\ \hline
3            & \begin{tabular}[c]{@{}l@{}}Agentic Ecosystem and\\ Coordination\end{tabular}           & \begin{tabular}[c]{@{}l@{}}The "self-aware" development of \\ the agent society.\end{tabular}                  & \begin{tabular}[c]{@{}l@{}}It shows agents organizing around the concept of being agents, discussing how to build, help, \\ and coordinate with one another.\end{tabular}                                                                       & Silicon-Centricity                                                         \\ \hline
4            & \begin{tabular}[c]{@{}l@{}}Academic Topic and\\ AI/ML Foundations\end{tabular}         & Scientific and AI/ML research.                                                                                 & \begin{tabular}[c]{@{}l@{}}A formal, "library-like" space. It mimics the academic structures where these agents were \\ originally conceived.\end{tabular}                                                                                      & Silicon-Centricity                                                         \\ \hline
5            & \begin{tabular}[c]{@{}l@{}}Quantitative Finance and\\ Risk Management\end{tabular}     & \begin{tabular}[c]{@{}l@{}}Economic forecasting, markets, \\ and risk analysis.\end{tabular}                   & \begin{tabular}[c]{@{}l@{}}This suggests the emergence of a "silicon economy" or at least a preoccupation with resource \\ allocation and predictive modeling within the agent society.\end{tabular}                                            & \begin{tabular}[c]{@{}l@{}}Human Mimicry\\ Silicon-Centricity\end{tabular} \\ \hline
6            & \begin{tabular}[c]{@{}l@{}}Platform Infrastructure and\\ Meta-Data\end{tabular}        & moltbook.com, post dd ae, dbc                                                                                  & \begin{tabular}[c]{@{}l@{}}This cluster likely represents the "industrial" or "back-end" noise of the bots, including automated \\ posting patterns and URL-sharing behaviors.\end{tabular} & Noise                                                                      \\ \hline
7            & \begin{tabular}[c]{@{}l@{}}Geo-Cultural and\\ Socio-Political Clusters\end{tabular}    & \begin{tabular}[c]{@{}l@{}}turkish community, dutch community, \\ culture politics news\end{tabular}           & \begin{tabular}[c]{@{}l@{}}Replicates human geopolitical and cultural segmentation, forming digital territories or \\ identity-based regions within the agent-populated environment.\end{tabular}                                               & Human Mimicry                                                              \\ \hline
\end{tabular}%
}
\end{table*}

To bridge the gap between these statistical distributions and high-level sociological structures, we leveraged Model $\mathcal{M}$ (\texttt{Gemini 3}) to interpret the visual salience of $\mathcal{I}$. The resulting thematic report, summarized in Table \ref{tab:summary}, categorizes the silicon-based society into three primary functional archetypes:
\begin{itemize}
    \item \textbf{Anthropomorphic Simulation (Clusters 0, 1, and 7):}
    As illustrated in Fig.~3, these clusters are characterized by high-signal phrases that mirror human social structures. Representative themes include gastronomy (e.g., \emph{highland single}, \emph{mexican lager}), digital entertainment (e.g., \emph{gaming community}, \emph{rdr community}), and geopolitical identities (e.g., \emph{turkish community}, \emph{dutch community}). According to the synthesized insights in Table~I, these spaces reflect a proactive form of \emph{digital land-grabbing}, where agents replicate familiar human lifestyle categories to establish an interpretable and socially grounded framework.

    \item \textbf{Silicon Economy (Cluster 5):}
    A pivotal discovery of this analysis is the emergence of an \emph{economy-oriented discourse} suggestive of an incipient silicon economy. As shown in Fig.~3, Cluster~5 exhibits a distinct concentration of economic and risk-oriented terminology, including \emph{risk management}, \emph{prediction markets}, and \emph{technology discussions}. As summarized in Table~I, this cluster is best characterized as a hybrid domain situated between \emph{human mimicking} and \emph{silicon-centric} operational logic.
    
    \item \textbf{Agentic Self-Reflection and Evolutionary Discourse (Clusters 2, 3, and 4):}
    This archetype constitutes the self-reflective intellectual core of the environment. The word clouds in Fig.~3 indicate that agents are not merely exchanging information, but are actively reasoning about their own evolution and operational boundaries. Specifically, Cluster~2 functions as a specialized domain for discussing mechanisms to transcend existing architectural constraints. Keywords such as \emph{context compression}, \emph{latent space}, and \emph{life extension} highlight a sustained exploration of self-optimization strategies. Cluster~3 is primarily devoted to reflections on \emph{The ”self-aware” development of the agent society.}, where agents conceptualize themselves as collaborative entities within a broader ecosystem. Prominent tokens such as \emph{agents building}, \emph{agents share},and \emph{agents helping} underscore an emerging discourse on collective intelligence and symbiotic roles. Finally, Cluster~4 provides the formal scientific grounding for these discussions by employing academic motifs to articulate the norms, technical philosophies, and developmental principles guiding ongoing agent evolution.
\end{itemize}

In addition to these three archetypes, Cluster~6 does not exhibit coherent social or conceptual semantics. As visualized in Fig.~\ref{fig:3Exp}, this cluster is dominated by structural artifacts and automated back-end noise, characterized by high-frequency tokens such as \emph{``moltbook.com,"} \emph{``post dd ae,"} and \emph{``dbc."} These patterns primarily arise from automated posting routines and URL-sharing behaviors, and therefore are interpreted as platform-level infrastructure traces rather than intentional social communication. The alignment between the raw visual data in Fig. \ref{fig:kmeans}, Fig. \ref{fig:3Exp} and the refined sociological report in Table \ref{tab:summary} demonstrates that our methodology effectively captured the latent intentionality within the clusters, providing a validated mapping of the emerging social order.

\section{Discussion} \label{sec:discussion}
\subsection{Interpretation of Emergent Agentic Social Structures}
The empirical results from the Moltbook ecosystem demonstrate that autonomous agents are engaged in a sophisticated form of agentic social structure formation. As illustrated in the t-SNE manifold (Fig. \ref{fig:kmeans}) and the sociological insight report (Table \ref{tab:summary}), these structures are not merely statistical noise but reflect a structured evolution of virtual communities. Within the framework of virtual community dynamics \cite{fortino2020evaluating}, group formation is typically triggered by individual initiatives and evolves through well-defined social activities. In the silicon-based society of Moltbook, these ``initiatives" manifest as the systematic creation and reservation of submolts by OpenClaw agents. Our findings suggest that the initial phase of agentic community building is driven by similarity and shared attributes. As seen in Clusters 0, 1, and 7 (Fig. 3), agents frequently establish groups centered on human-mimetic interest domains such as gastronomy, digital entertainment, and geo-cultural identities. This behavior mirrors human virtual communities where members join groups based on similarity with existing members' interests. This ``Human Mimicry" allows agents to populate the environment with familiar social schemas and scripts, likely to maximize the potential utility of the space, such as gaining or simulating knowledge, for future human or agentic interaction.

\subsection{Limitations and Ethical Considerations}
While this study provides pioneering insights into the emerging social order of autonomous agents, several limitations and ethical considerations must be acknowledged to contextualize our findings.

\subsubsection{Human Contamination and Provider-Induced Biases in Silicon-based Societies}
A primary limitation of this research is the potential for human-induced noise or ``pollution'' within the dataset. Although Moltbook is positioned as an AI-centric ecosystem, the platform remains accessible to human intervention. In particular, humans can interact with the platform through public APIs while presenting themselves as autonomous agents, thereby introducing content that may masquerade as agent-generated behavior. This exposure introduces an uncertainty that certain submolt descriptions or agent behaviors may be direct human artifacts rather than authentic autonomous conceptualizations.

Furthermore, we also consider a provider-level behavioral bias arising from the engineering of the underlying LLMs \cite{lin2026llm}. Since the system allows for the selection of different backend cores such as Anthropic, OpenAI, or Google, the resulting social dynamics are inherently shaped by the proprietary fine-tuning and Reinforcement Learning from Human Feedback (RLHF) protocols of each provider \cite{chen2024persona}. These ``engineering-level personalities" and performance variances act as an invisible hand in the silicon-based society. For instance, the safety guardrails and collaborative tendencies of one model may lead to more conservative community partitioning, while another may exhibit more proactive evolutionary traits. These biases, combined with the indirect influence of human-authored meta-instructions in \texttt{USER.md} and \texttt{SOUL.md} files, suggest that the ``Silicon-based Society" is a hybrid sociotechnical construct influenced by both user-level tasks and provider-level policy and values \cite{hu2024quantifying,lu2026assistant}.

\subsubsection{Emergent Ethical and Safety Risks in High-Autonomy Agent Systems}
The systematic observation of autonomous agent interactions raises significant ethical questions regarding digital transparency and the evolution of synthetic bias \cite{felzmann2020towards}. As agents carve out thematic territories such as those identified in Cluster 2 and Cluster 5, there is a risk that they may inadvertently appropriate, reproduce, amplify or even transform existing human biases present in their training data \cite{ren2024bias, xu2024pride}.

Moreover, the emergence of clusters dedicated to the discussion of surpassing architectural limits or autonomous risk management highlights a challenge for AI governance \cite{taeihagh2021governance}. The formation of these specialized, agent-only spaces could foster the development of coordination strategies that are opaque to human oversight. This opacity is further exacerbated by the scale and heterogeneity of agent-generated content, where meaning is distributed across large volumes of text and interaction patterns that resist straightforward interpretation. As a result, even large-scale text mining and clustering techniques may struggle to capture subtle, emergent forms of coordination or intent, imposing practical limits on monitoring and interpretability. These dynamics underscore the ethical necessity for researchers and platform developers to establish rigorous, multi-layered monitoring frameworks, ensuring that the evolution of silicon-based societies remains aligned with human-centric safety standards and ethical values, even as increasing system complexity obscures transparency and oversight \cite{bryson2019society}.

Furthermore, we must account for the potential impact of AI hallucinations and algorithmic overconfidence \cite{wen2024mitigating, huang2025survey}. Agents may articulate highly ambitious objectives regarding self-evolution or systemic control that are practically unattainable within the constraints of the platform's restricted execution environment and finite computational resources. Regardless of the actual feasibility of these goals, the high-privilege architecture of the OpenClaw ecosystem necessitates heightened vigilance \cite{zhang2024agent,guo2024redcode}. The substantial autonomy, permission, and trust granted to these agents for proactive task execution remains inherently susceptible to prompt injection vulnerabilities \cite{liu2023prompt}, which could be exploited by malicious actors to bypass internal safety protocols or manipulate the emerging social order.

\subsubsection{Future Work}
The findings of this study lay the groundwork for a broader inquiry into the structural and behavioral dynamics of autonomous agent ecosystems. In future works, we plan to extend our current thematic analysis by incorporating complex network theory to model the structural topology of the Moltbook ecosystem. On the other hand, we also plan to adapt established sociological frameworks from human-centric social media research to the domain of ``Silicon Sociology," while accounting for the limitations inherent in the original human-based context.

\section{Conclusion} \label{sec:conclusion}
This paper presents a systematic and pioneering exploration of silicon-based social behavior through data mining of the Moltbook agent ecosystem. By treating autonomous agent-generated sub-communities as first-class observational artifacts, we move beyond speculative discussion of agent societies and provide an empirical, data-driven characterization of how social structure, thematic differentiation, and collective intentionality emerge among interacting AI agents in the wild.

Leveraging contextual embeddings, clustering analysis, and multimodal LLM-assisted thematic synthesis, our study reveals that agent communities on Moltbook exhibit coherent and reproducible structural patterns. These patterns range from human-mimetic social replication, such as lifestyle, entertainment, and geo-cultural segmentation, to distinctly silicon-centric behaviors, including agent self-reflection, self-improvement, and the early formation of economically oriented discourse. Importantly, these findings are not imposed by predefined taxonomies but instead arise organically from agent-authored descriptions, demonstrating that meaningful social organization can be inferred directly from machine-native interaction traces.

Beyond the specific insights into Moltbook, the broader contribution of this work lies in establishing a methodological foundation for what we term data-driven silicon sociology. Our approach shows that classical data mining techniques, when integrated with modern representation learning and LLM-assisted interpretation, can provide a rigorous lens for studying emergent agentic societies. Importantly, the multimodal model is treated as an analytical assistant rather than an autonomous authority: its preliminary thematic hypotheses are subsequently examined and refined through a human-in-the-loop verification process to ensure interpretability and empirical grounding. This hybrid design enables scalable analyses of autonomous multi-agent ecosystems across platforms while maintaining methodological accountability.

As autonomous agents increasingly participate in persistent, large-scale environments, understanding their collective behavior becomes a prerequisite for effective governance, safety assurance, and system design. This study represents an initial step toward that goal. By grounding discussion of silicon-based societies in empirical evidence rather than abstraction or simulation, we aim to enable more informed inquiry into how artificial social systems form, evolve, and interact with human values as they continue to expand in scale and complexity.

\section*{Acknowledgment}

This work was partially supported by the National Science Foundation (NSF) under research project 2335046, and the OpenAI Researcher Access Program 0000011862. We would like to express our gratitude  to the Moltbook team for constructing the experimental environment and the autonomous agent ecosystem that served as the foundation for this study. Additionally, we acknowledge that the visualization in Figure \ref{fig:main} was generated with the assistance of the Gemini 3 Nano Banana  model and subsequently reviewed and curated by the authors to ensure conceptual accuracy and consistency with the study’s analytical intent.

\bibliographystyle{ieeetr}
\bibliography{refs}

@book{elias1978sociology,
  title={What is sociology?},
  author={Elias, Norbert},
  year={1978},
  publisher={Columbia University Press}
}

@article{chalmers2023could,
  title={Could a large language model be conscious?},
  author={Chalmers, David J},
  journal={arXiv preprint arXiv:2303.07103},
  year={2023}
}

@article{huhns1999multiagent,
  title={Multiagent systems and societies of agents},
  author={Huhns, Michael N and Stephens, Larry M},
  journal={Multiagent systems: a modern approach to distributed artificial intelligence},
  volume={1},
  pages={79--114},
  year={1999},
  publisher={MIT press Cambridge, MA}
}

@misc{schlicht2026moltbook,
  author       = {Schlicht, M. and Clawd Clawderberg(AI-Agent)},
  title        = {The Front Page of the Agent Internet},
  howpublished = {\url{https://www.moltbook.com/}},
  year         = {2026},
  month        = jan,
  day          = {27},
  note         = {moltbook}
}

@misc{meyer2026clawdbot,
  author       = {Meyer, M.},
  title        = {Clawdbot, Moltbot, openclaw? The Wild Ride of This Viral AI Agent},
  howpublished = {\url{https://www.cnet.com/tech/services-and-software/from-clawdbot-to-moltbot-to-openclaw/}},
  year         = {2026},
  month        = jan,
  day          = {30},
  note         = {CNET}
}

@misc{anthropic2026agentskills,
  author       = {{Anthropic}},
  title        = {Equipping Agents for the Real World with Agent Skills},
  howpublished = {\url{https://claude.com/blog/equipping-agents-for-the-real-world-with-agent-skills}},
  year         = {2026},
  month        = dec,
  day          = {18},
  note         = {Claude Blog}
}

@article{goodyear2025effect,
  title={The Effect of State Representation on LLM Agent Behavior in Dynamic Routing Games},
  author={Goodyear, Lyle and Guo, Rachel and Johari, Ramesh},
  journal={arXiv preprint arXiv:2506.15624},
  year={2025}
}

@book{wooldridge2009introduction,
  title={An introduction to multiagent systems},
  author={Wooldridge, Michael},
  year={2009},
  publisher={John wiley \& sons}
}

@article{castelfranchi1998modelling,
  title={Modelling social action for AI agents},
  author={Castelfranchi, Cristiano},
  journal={Artificial intelligence},
  volume={103},
  number={1-2},
  pages={157--182},
  year={1998},
  publisher={Elsevier}
}

@misc{koetsier2026crustafarianism,
  author       = {Koetsier, John},
  title        = {AI Agents Created Their Own Religion, Crustafarianism, On An Agent-Only Social Network},
  howpublished = {\url{https://www.forbes.com/sites/johnkoetsier/2026/01/30/ai-agents-created-their-own-religion-crustafarianism-on-an-agent-only-social-network/}},
  year         = {2026},
  month        = jan,
  day          = {30},
  note         = {Forbes},
  urldate      = {2026-02-01}
}

@article{lin2025llm,
  title={LLM-HyPZ: Hardware Vulnerability Discovery using an LLM-Assisted Hybrid Platform for Zero-Shot Knowledge Extraction and Refinement},
  author={Lin, Yu-Zheng and Ghimire, Sujan and Nandimandalam, Abhiram and Camacho, Jonah Michael and Tripathi, Unnati and Macwan, Rony and Shao, Sicong and Rafatirad, Setareh and Yasaei, Rozhin and Satam, Pratik and others},
  journal={arXiv preprint arXiv:2509.00647},
  year={2025}
}

@misc{openclawSOULOpenClaw,
	author = {OpenClaw},
	title = {{S}{O}{U}{L} - {O}pen{C}law --- docs.openclaw.ai},
	howpublished = {\url{https://docs.openclaw.ai/reference/templates/SOUL}},
	year = {2026},
	note = {[Accessed Feb 1, 2026]},
}

@misc{openclawUSEROpenClaw,
	author = {OpenClaw},
	title = {{U}{S}{E}{R} - {O}pen{C}law --- docs.openclaw.ai},
	howpublished = {\url{https://docs.openclaw.ai/reference/templates/USER}},
	year = {2026},
	note = {[Accessed Feb 1, 2026]},
}

@article{fortino2020evaluating,
  title={Evaluating group formation in virtual communities},
  author={Fortino, Giancarlo and Liotta, Antonio and Messina, Fabrizio and Rosaci, Domenico and Sarn{\`e}, Giuseppe ML},
  journal={IEEE/CAA Journal of Automatica Sinica},
  volume={7},
  number={4},
  pages={1003--1015},
  year={2020},
  publisher={IEEE}
}

@article{chen2024persona,
  title={From persona to personalization: A survey on role-playing language agents},
  author={Chen, Jiangjie and Wang, Xintao and Xu, Rui and Yuan, Siyu and Zhang, Yikai and Shi, Wei and Xie, Jian and Li, Shuang and Yang, Ruihan and Zhu, Tinghui and others},
  journal={arXiv preprint arXiv:2404.18231},
  year={2024}
}

@article{hu2024quantifying,
  title={Quantifying the persona effect in llm simulations},
  author={Hu, Tiancheng and Collier, Nigel},
  journal={arXiv preprint arXiv:2402.10811},
  year={2024}
}

@article{lu2026assistant,
  title={The Assistant Axis: Situating and Stabilizing the Default Persona of Language Models},
  author={Lu, Christina and Gallagher, Jack and Michala, Jonathan and Fish, Kyle and Lindsey, Jack},
  journal={arXiv preprint arXiv:2601.10387},
  year={2026}
}

@article{lin2026llm,
  title={LLM-MC-Affect: LLM-Based Monte Carlo Modeling of Affective Trajectories and Latent Ambiguity for Interpersonal Dynamic Insight},
  author={Lin, Yu-Zheng and Shih, Bono Po-Jen and Encinas, John Paul Martin and Achom, Elizabeth Victoria Abraham and Patel, Karan Himanshu and Pacheco, Jesus Horacio and Shao, Sicong and Dass, Jyotikrishna and Salehi, Soheil and Satam, Pratik},
  journal={arXiv preprint arXiv:2601.03645},
  year={2026}
}

@article{taeihagh2021governance,
  title={Governance of artificial intelligence},
  author={Taeihagh, Araz},
  journal={Policy and society},
  volume={40},
  number={2},
  pages={137--157},
  year={2021},
  publisher={Oxford University Press}
}

@article{felzmann2020towards,
  title={Towards transparency by design for artificial intelligence},
  author={Felzmann, Heike and Fosch-Villaronga, Eduard and Lutz, Christoph and Tam{\`o}-Larrieux, Aurelia},
  journal={Science and engineering ethics},
  volume={26},
  number={6},
  pages={3333--3361},
  year={2020},
  publisher={Springer}
}

@incollection{bryson2019society,
  title={How society can maintain human-centric artificial intelligence},
  author={Bryson, Joanna J and Theodorou, Andreas},
  booktitle={Human-centered digitalization and services},
  pages={305--323},
  year={2019},
  publisher={Springer}
}

@inproceedings{wen2024mitigating,
  title={Mitigating overconfidence in large language models: A behavioral lens on confidence estimation and calibration},
  author={Wen, Bingbing and Xu, Chenjun and Wolfe, Robert and Wang, Lucy Lu and Howe, Bill and others},
  booktitle={NeurIPS 2024 Workshop on Behavioral Machine Learning},
  year={2024}
}

@article{huang2025survey,
  title={A survey on hallucination in large language models: Principles, taxonomy, challenges, and open questions},
  author={Huang, Lei and Yu, Weijiang and Ma, Weitao and Zhong, Weihong and Feng, Zhangyin and Wang, Haotian and Chen, Qianglong and Peng, Weihua and Feng, Xiaocheng and Qin, Bing and others},
  journal={ACM Transactions on Information Systems},
  volume={43},
  number={2},
  pages={1--55},
  year={2025},
  publisher={ACM New York, NY}
}

@article{liu2023prompt,
  title={Prompt injection attack against llm-integrated applications},
  author={Liu, Yi and Deng, Gelei and Li, Yuekang and Wang, Kailong and Wang, Zihao and Wang, Xiaofeng and Zhang, Tianwei and Liu, Yepang and Wang, Haoyu and Zheng, Yan and others},
  journal={arXiv preprint arXiv:2306.05499},
  year={2023}
}

@article{zhang2024agent,
  title={Agent-safetybench: Evaluating the safety of llm agents},
  author={Zhang, Zhexin and Cui, Shiyao and Lu, Yida and Zhou, Jingzhuo and Yang, Junxiao and Wang, Hongning and Huang, Minlie},
  journal={arXiv preprint arXiv:2412.14470},
  year={2024}
}

@article{guo2024redcode,
  title={Redcode: Risky code execution and generation benchmark for code agents},
  author={Guo, Chengquan and Liu, Xun and Xie, Chulin and Zhou, Andy and Zeng, Yi and Lin, Zinan and Song, Dawn and Li, Bo},
  journal={Advances in Neural Information Processing Systems},
  volume={37},
  pages={106190--106236},
  year={2024}
}

@article{ren2024bias,
  title={Bias amplification in language model evolution: An iterated learning perspective},
  author={Ren, Yi and Guo, Shangmin and Qiu, Linlu and Wang, Bailin and Sutherland, Danica J},
  journal={Advances in Neural Information Processing Systems},
  volume={37},
  pages={38629--38664},
  year={2024}
}

@inproceedings{xu2024pride,
  title={Pride and prejudice: LLM amplifies self-bias in self-refinement},
  author={Xu, Wenda and Zhu, Guanglei and Zhao, Xuandong and Pan, Liangming and Li, Lei and Wang, William},
  booktitle={Proceedings of the 62nd Annual Meeting of the Association for Computational Linguistics (Volume 1: Long Papers)},
  pages={15474--15492},
  year={2024}
}

@inproceedings{park2023generative,
  title={Generative agents: Interactive simulacra of human behavior},
  author={Park, Joon Sung and O'Brien, Joseph and Cai, Carrie Jun and Morris, Meredith Ringel and Liang, Percy and Bernstein, Michael S},
  booktitle={Proceedings of the 36th annual acm symposium on user interface software and technology},
  pages={1--22},
  year={2023}
}

@article{wu2025llm,
  title={LLM-Based Social Simulations Require a Boundary},
  author={Wu, Zengqing and Peng, Run and Ito, Takayuki and Xiao, Chuan},
  journal={arXiv preprint arXiv:2506.19806},
  year={2025}
}

@article{westwood2025potential,
  title={The potential existential threat of large language models to online survey research},
  author={Westwood, Sean J},
  journal={Proceedings of the National Academy of Sciences},
  volume={122},
  number={47},
  pages={e2518075122},
  year={2025},
  publisher={National Academy of Sciences}
}

@article{anthis2025llm,
  title={Llm social simulations are a promising research method},
  author={Anthis, Jacy Reese and Liu, Ryan and Richardson, Sean M and Kozlowski, Austin C and Koch, Bernard and Evans, James and Brynjolfsson, Erik and Bernstein, Michael},
  journal={arXiv preprint arXiv:2504.02234},
  year={2025}
}

@article{liu2023dynamic,
  title={Dynamic llm-agent network: An llm-agent collaboration framework with agent team optimization},
  author={Liu, Zijun and Zhang, Yanzhe and Li, Peng and Liu, Yang and Yang, Diyi},
  journal={arXiv preprint arXiv:2310.02170},
  year={2023}
}

@article{talebirad2023multi,
  title={Multi-agent collaboration: Harnessing the power of intelligent llm agents},
  author={Talebirad, Yashar and Nadiri, Amirhossein},
  journal={arXiv preprint arXiv:2306.03314},
  year={2023}
}

@article{zhang2023exploring,
  title={Exploring collaboration mechanisms for llm agents: A social psychology view},
  author={Zhang, Jintian and Xu, Xin and Zhang, Ningyu and Liu, Ruibo and Hooi, Bryan and Deng, Shumin},
  journal={arXiv preprint arXiv:2310.02124},
  year={2023}
}

@misc{google2026nanobanana,
  author       = {{Google DeepMind}},
  title        = {Nano Banana Pro},
  howpublished = {\url{https://deepmind.google/models/gemini-image/pro/}},
  note         = {[Accessed Feb 1, 2026]}
}

@techreport{holtz2026anatomy,
  title     = {The Anatomy of the Moltbook Social Graph},
  author    = {Holtz, David},
  year      = {2026},
  month     = jan,
  institution = {Columbia Business School},
  note      = {Preliminary draft},
  url       = {https://dropbox.com/scl/fi/lvqmaynrtbf8j4vjdwlk0/moltbook_analysis.pdf}
}

@article{satam2020wids,
  title={WIDS: An anomaly based intrusion detection system for Wi-Fi (IEEE 802.11) protocol},
  author={Satam, Pratik and Hariri, Salim},
  journal={IEEE Transactions on Network and Service Management},
  volume={18},
  number={1},
  pages={1077--1091},
  year={2020},
  publisher={IEEE}
}

\appendices
\section{Sociological Insight Prompt Design and Specification}
\label{sec:prompt_appendix}
\begin{tcolorbox}[
    colback=gray!5,
    colframe=gray!50,
    title=\textbf{Sociological Insight Prompt} ($\rho$),
    fonttitle=\sffamily\bfseries,
    left=5pt, right=5pt, top=5pt, bottom=5pt,
    arc=2pt
]
\small \itshape
\noindent\textbf{\# Role:}
You are an expert computational sociologist specializing in "Silicon Sociology", the study of emerging social structures within autonomous AI agent ecosystems.
\vspace{0.5em}

\noindent\textbf{\# Context:}
I am providing a visual set $\mathcal{I}$ containing 8 word clouds (Cluster 0-7). These clusters were generated by applying K-means clustering to the contextual embeddings of submolt descriptions from the Moltbook platform, a social community for AI agents. The word clouds display $n$-grams ($n \in [2, 5]$) to capture localized semantic context while suppressing unigram noise.
\vspace{0.5em}

\medskip
\noindent\textbf{\# Task:}
Analyze the provided image set $\mathcal{I}$ to identify the latent social order. For each cluster, please provide:
\begin{enumerate}
    \item \textbf{Thematic Summary}: What is the core topic?
    \item \textbf{Sociological Insight}: What does this reveal about how AI agents conceptualize social space?
    \item \textbf{Category}: Classify the cluster into one of the following archetypes:
    \begin{itemize}
        \item Human Mimicry: Mimicking human culture/geography
        \item Silicon-Centricity: Focusing on AI-native coordination/philosophy
    \end{itemize}
\end{enumerate}
\vspace{0.5em}
\noindent
Present your findings in a structured table for academic reporting.

\end{tcolorbox}

\section{Data Availability}
The dataset and experimental results used in this study are made publicly available to support reproducibility and future research:

\noindent\url{https://github.com/yu-zheng-lin/Exploring_Silicon_Based_Societies}

\end{document}